\documentclass{article}

\usepackage{microtype}
\usepackage{graphicx}
\usepackage{subfigure}
\usepackage{booktabs} 

\usepackage{hyperref}

\usepackage{tikz}
\usetikzlibrary{positioning,arrows,automata,calc}

\usepackage[accepted]{icml2025}

\usepackage{amsmath}
\usepackage{amssymb}
\usepackage{mathtools}
\usepackage{amsthm}
\usepackage{enumitem}

\usepackage[capitalize,noabbrev]{cleveref}

\theoremstyle{plain}

\theoremstyle{definition}

\theoremstyle{remark}

\usepackage[textsize=tiny]{todonotes}

\icmltitlerunning{Authenticated Delegation}

\begin{document}

\twocolumn[
\icmltitle{Authenticated Delegation and Authorized AI Agents}



\icmlsetsymbol{equal}{*}

\begin{icmlauthorlist}

\icmlauthor{Tobin South}{mit}
\icmlauthor{Samuele Marro}{oxford}
\icmlauthor{Thomas Hardjono}{mit}
\icmlauthor{Robert Mahari}{mit,hls}
\icmlauthor{Cedric Deslandes Whitney}{berkley}
\icmlauthor{Dazza Greenwood}{mit}
\icmlauthor{Alan Chan}{govai}
\icmlauthor{Alex Pentland}{mit,stanford}
\end{icmlauthorlist}

\icmlaffiliation{mit}{MIT, Cambridge, MA, USA}
\icmlaffiliation{hls}{Harvard Law School, Cambridge, MA, USA}
\icmlaffiliation{berkley}{University of California, Berkeley
Berkeley, California, USA}
\icmlaffiliation{oxford}{Department of Engineering Science, University of Oxford, UK}
\icmlaffiliation{govai}{Centre for the Governance of AI, Oxford, UK}
\icmlaffiliation{stanford}{Stanford University, Palo Alto, CA, USA}

\icmlcorrespondingauthor{Tobin South}{tsouth@mit.edu}

\icmlkeywords{Machine Learning, ICML}

\vskip 0.3in
]



\printAffiliationsAndNotice{}  

\begin{abstract}
The rapid deployment of autonomous AI agents creates urgent challenges around authorization, accountability, and access control in digital spaces. 
New standards are needed to know whom AI agents act on behalf of and guide their use appropriately, protecting online spaces while unlocking the value of task delegation to autonomous agents. 
We introduce a novel framework for authenticated, authorized, and auditable delegation of authority to AI agents, where human users can securely delegate and restrict the permissions and scope of agents while maintaining clear chains of accountability.
This framework builds on existing identification and access management protocols, extending OAuth 2.0 and OpenID Connect with agent-specific credentials and metadata, maintaining compatibility with established authentication and web infrastructure.
Further, we propose a framework for translating flexible, natural language permissions into auditable access control configurations, enabling robust scoping of AI agent capabilities across diverse interaction modalities.
Taken together, this practical approach facilitates immediate deployment of AI agents while addressing key security and accountability concerns, working toward ensuring agentic AI systems perform only appropriate actions and providing a tool for digital service providers to enable AI agent interactions without risking harm from scalable interaction. 
\end{abstract}

\section{Introduction}

Agentic AI systems, also referred to as AI assistants or simply `agents', are AI systems that can pursue complex goals with limited direct supervision on behalf of a user \cite{Gabriel2024-cm,chan2024visibility,shavit2023practices,chan2023harms,Kenton2023-cz}, including by interacting with a variety of external digital tools and services \cite{Nakano2021-uj,Lieberman1997-qm,fourney2024magentic}. For example, AI agents given a prompt to book travel arrangements for a holiday may browse the web for recommendations, search for flights via APIs, or message an airline agent in natural language via chat services to arrange a booking. Such communications could even extend to AI agent negotiations \cite{Abdelnabi2023-al} and other multi-agent contexts. 

While current AI agents have limitations \cite{Raji2022-kj,Wang2023-rt}, lack the ability to perform certain tasks \cite{Liu2023-rd}, and may be susceptible to attacks such as prompt injections \cite{Yao2024-uv,Liu2023-bl,Zhu2023-yf}, there has been rapid progress in their development and commercial interest. 

This has raised many concerns over the risks of AI agents and how they should be governed \cite{shavit2023practices,Gabriel2024-cm,eu_ai_act}. 
Credentials and verification may become critical in verifying the properties and metadata of AI systems \cite{chan2024visibility},
uniquely identifying humans in online spaces \cite{Borge2017-ti} (or at least distinguishing humans from AI agents \cite{adler2024personhood}), protecting contextual confidence \cite{Jain2023-ij}, mitigating AI-augmented influence operations \cite{Goldstein2023-co}, preventing AI manipulation of humans \cite{Bai2023-vc,Singh2024-dz}, and governing or auditing AI systems more broadly \cite{reuel2024open,South2023-wh}. 
The world needs ways to explicitly delegate authority to agents, transparently identify those agents as AI, and enforce human-centered choices around security and permission for these agents.

We distinguish three key concepts: \textit{authentication} confirms an entity’s identity; \textit{authorization} determines the permissible actions and resource accesses that the authenticated identity is allowed to perform, defining the scope and limitations of delegated activity; and \textit{auditability} allows all parties to inspect and verify that claims, credentials, and attributes remain unaltered, supporting trustworthy authentication and authorization decisions.

This work has three key contributions. 
First, \autoref{sec:why} builds upon the existing literature to outline \textbf{why authenticated delegation is important} for AI agents, and what risks it could mitigate. In doing so, we provide an overview of current practices and where they fall short.
Second, \autoref{sec:technical} directly addresses this need by \textbf{extending existing authentication and authorization protocols to enable authenticated delegation} to AI agents, examining the role OpenID Connect and OAuth 2.0 could play in enabling a pragmatic, robust, and extensible implementation.
Third, \autoref{sec:scoping} explores the \textbf{role of agentic access control} and 
outlines a method for \textbf{expressing flexible, natural language permissions for agents} and transforming them into auditable, fine-grained access control rules, that can operate across agent modalities (e.g., web requests, computer use, or language interfaces),
Further, this work provides \textbf{example use cases} of the framework in \autoref{sec:examples} and a \textbf{legal analysis of the implications} of this work in \autoref{sec:legal}.

\begin{figure}
	\centering
	\includegraphics[width=\linewidth]{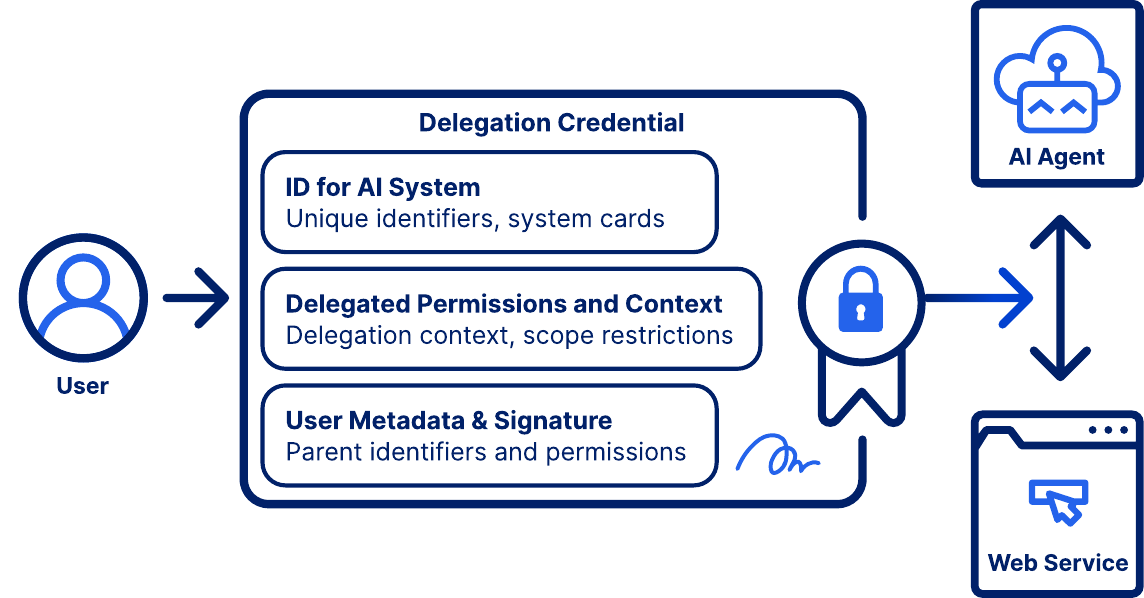}
	\caption{Conceptual overview of a verifiable delegation credential for AI agents. Users issue delegation credentials that include: the AI system's unique identity and properties, delegated permissions with contextual scope restrictions, user metadata, and cryptographic signatures for verifiability. These credentials enable secure, trustworthy interactions between AI agents and third-party services, ensuring traceability and appropriate delegation of authority. }
\end{figure}

\section{Why authenticated delegation is important}\label{sec:why}
Authenticated delegation is the process of instructing an AI system to perform a task that requires access to tools, the web, or computer environments in such a way that third parties can verify that (a) the interacting entity is an AI agent, (b) that the AI agent is acting on behalf of a specific human user, and (c) that the AI agent has been granted the necessary permissions to perform specific actions. 

Verifying the properties of interacting entities will be relevant whenever a context exists where an AI agent \emph{could} act on behalf of a human user, and especially where the agent is capable of taking consequential actions. This remains true whether the AI system is run locally or provided by an AI vendor---as harm can occur in both---and must be able to operate across various digital contexts and with AI models of heterogeneous capabilities.

At a high level, authenticated delegation involves a human user creating a digital authorization that a specific AI agent can use to access a digital service (or interact with another AI agent) on behalf of the user, which can be verified by the corresponding service or agent for its authenticity. Such authorization can include additional information, such as unique identifiers for the agent instance, permissions on what the agent is allowed to do, and other information (e.g., the capabilities and failure modes of the agent or information about the human user).
The authorization must be uniquely and cryptographically linked to the digital identity of the human delegator who granted the authorization. 
This could be done by linking to email accounts (as is commonplace for application accounts), linking to a more robust digital identity, or via domain-specific identifiers (such as user accounts within an organizational setting).

In practice, this needn't be substantially different from existing authentication and authorization mechanisms used today, such as how a calendar application is authorized to access a user's calendar data and scan it for upcoming events. However, AI agents' autonomous and highly capable nature means more care is needed in how we manage delegation. As such, let us examine the use cases for authenticated delegation in more detail.

\subsection{Functions of authenticated delegation}
Authenticated delegation opens avenues for AI agents to accelerate complex tasks, automate workflows, and seamlessly interface with digital services on behalf of human users. However, granting such agency also entails risks around scope misalignment, resource abuse, or a breakdown of clear accountability. This subsection delineates how robust identity verification, explicit scoping, and mutual authentication can unleash practical use cases—ranging from streamlined enterprise processes to safe, multi-agent coordination—while mitigating key vulnerabilities. By highlighting both the opportunities and the potential pitfalls, we underscore why adopting secure, verifiable delegation mechanisms is vital to responsibly harness AI agents.

\subsubsection{Current challenges in delegating authority to AI agents}
As the capabilities of LLMs improve, there is a growing interest in making them more autonomous and general-purpose. A key aspect of this is the ability to use tools or access external services. For simple tasks such as asking an agent to search the web for information, write and execute code, or generate an image, this is straightforward and does not require additional authorization or individual-specific security mechanisms. However, to unlock use cases such as interacting with personal or organizational accounts, accessing sensitive personal information, or interacting with consequential infrastructure, more robust delegation frameworks are needed.

\textbf{Example:} Consider the above example of an AI agent booking a holiday. Having an agent search the web for information may not need any authorization, but how could that agent access a user's calendar or make a purchase? 
For calendars, users are used to the expected flow of granting access to applications to access their calendar data. This would be no different for an AI agent (and would be naively supported in the solution outlined in \autoref{sec:technical})--indeed, limited OAuth~2.0 support is enabled in some agent tools such as OpenAI GPT actions. 
Now consider a flight purchase. You \emph{could} provide your credit card details in the context window for the agent and prompt it to follow at budget, but this introduces a variety of security concerns and relies upon the underlying reliability of the AI system to not take unexpected actions or be vulnerable to attacks or jailbreaks. 
Instead, an AI agent should be authenticated and authorized to make a purchase on specific booking services, where credit cards are stored securely, and where explicit spending limits can be enforced.

\subsubsection{Communicating limitations and restricting scope}
Current approaches to limiting the scope of AI agents are limited and one-sided. A user can provide a strong prompt to an agent to limit its actions, but this comes with a variety of failure modes \citep{Liu2023-bl}. Access to tools or websites can be blocked, but this is limited in the granularity of control. An AI system deployer could implement further controls, such as monitoring and blocking specific actions or website subdomains when agentic functionality occurs, but doesn't communicate these limitations to the service the agent is interacting with. 
By explicitly limiting the scope of an AI agent and communicating these limitations to the service the agent is interacting with, we can enable a more robust interaction between AI agents and services. A more detailed examination of how this could be designed across web, API, and natural language access modalities is available in \autoref{sec:scoping}.

\textbf{Example:} An AI agent is used by a physician to provide diagnostic recommendations in a telemedicine portal, logging in with basic credentials that do not specify its limitations. The portal assumes full physician capabilities, granting the agent access to all patient records, including a video assessment with a voice recording from a specialist. The agent, being text-only and unable to process video, generates a diagnosis based solely on the text data, appending a standard caveat—``not all available information was used''—which is overlooked. Trusting the incomplete recommendation, the physician risks making a misinformed treatment decision. If the agent's limitations were explicitly communicated via authenticated delegation, the portal could have flagged the need for a human review of the multimedia content, avoiding a potentially harmful oversight.

\subsubsection{Verification in multi-agent communication}
When AI agents communicate to collaborate on tasks or facilitate interactions, ensuring mutual authentication becomes paramount. 
Securing communication channels is not enough; agents must also verify that they authentically represent the users or organizations they claim to represent. Mutual authentication ensures that agents can trust each other's intentions, capabilities, and authority, preventing impersonation, unauthorized actions, and potential misuse. This verification is essential for fostering reliable, safe, and accountable multi-agent ecosystems.

\textbf{Example:} Two AI agents—one representing a hospital and the other an insurance company—collaborate to process a patient's claim. The hospital agent submits treatment details, while the insurance agent verifies coverage. Without mutual authentication, a third-party malicious agent could impersonate the hospital, submitting fraudulent claims, or the insurance agent could reject valid claims out of concern over authenticity. 

\subsubsection{Protecting human spaces online}
As AI agents grow increasingly adept at mimicking human behavior---crafting text, creating personas, and even replicating nuanced human interactions---it becomes harder to maintain digital environments genuinely inhabited by real people. This challenge drives the need for safe, human-only online spaces where authenticity is preserved and scalable manipulation is curbed by verifying human personhood \cite{adler2024personhood}. However, many AI agents act as useful proxies, assistants, or representatives for human users who cannot, or prefer not to, engage directly. Authenticated delegation enables these spaces to be selectively accessible to AI agents, while still ensuring that the AI agents are linked to verified human principals.

\begin{figure}[h]
	\centering
	\includegraphics[width=0.8\linewidth]{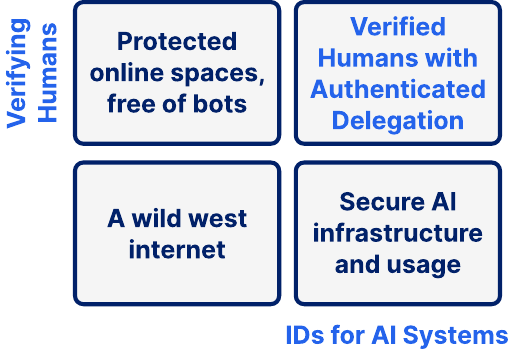}
	\caption{Authenticated delegation can benefit from user identification or verification of personhood (e.g., through personhood credentials). By combining verified human identity with authenticated delegation, we can support safer online spaces for human interaction while enabling the trustworthy and controlled use of AI agents.}
\end{figure}

\textbf{Example:} The Australian government's recent social media ban demonstrates how governments can restrict access to online spaces by requiring users to prove their age, often through methods like a government-issued ID or a face scan. While these measures aim to limit underage access, they may also inadvertently block AI agents from accessing these platforms. Instead of a blanket restriction, platforms could explicitly allow AI agents to access their services in controlled ways by leveraging authenticated delegation. This approach would ensure that AI agents act transparently on behalf of verified human users. For instance, an agent could access a user's social media account to retrieve information about friends and help draft an email, all while maintaining compliance with platform policies and ensuring accountability.

\subsubsection{Supporting contextual integrity}
Contextual integrity addresses adherence to context-specific norms and privacy, which include actors (who is involved in the information flow), attributes (what information is shared), transmission principles (under what conditions information is shared), and social context (the broader cultural, institutional, or situational environment shaping these norms) \cite{ghalebikesabie2024,zhan2022,fan2024,nissenbaum2004privacy}. 
Contextual integrity offers a perspective for reasoning about how AI agents can act in ways that are contextually appropriate, transparent, and aligned with societal norms and the expectations of their human delegators \cite{bagdasarian2024,ghalebikesabie2024,bloom2022}. 
This includes exploring which decisions can reasonably be made autonomously by the AI and under what conditions human oversight or intervention might be necessary (e.g., when is human-in-the-loop required and who is responsible). 

\textbf{Example:} 
An AI assistant with authenticated delegation can be issued distinct credentials for separate contexts  (e.g., an enterprise-context assistant and a personal one).
Each credential encodes the agent’s information, the delegating user, and context-specific permissions. 
By enforcing these scoped credentials, services can ensure that the assistant adheres to contextual integrity and rejects actions that cross boundaries, such as using information from work documents to complete personal forms. 
This separation of roles and explicit permission-sharing protects privacy, ensures accountability, and safeguards human oversight for cross-context decisions.

\subsection{Background}
Authenticated delegation can address various challenges, from traceability of AI outcomes to limitations on what spaces can be accessed and actions taken by AI systems. The overarching aim of identification and credentialing systems is to facilitate secure online environments and authenticated access to services. To this end, a variety of existing protocols and standards have been developed, tailored to both human users and AI systems, to uphold these goals in different contexts.

\paragraph{Comparisons to other AI identifiers}
To \emph{verify human identity online}, a large body of work exists ranging from simple authentication such as {OAuth~2.0}~\cite{RFC6749-Formatted} to more complex digital identity frameworks as W3C's Verifiable Credentials \citep{noauthor_verifiable_2024}, decentralized identifiers \citep{sporny_decentralized_2024}, and the European Union Digital Identity's privacy-preserving digital wallets \citep{eudi_arf_2024}. 
To \emph{privately prove personhood}, a number of systems have been developed to distinguish human users from bots, including proof-of-personhood systems designed to counter automated Sybil attacks \citep{borge_proof_personhood_2017}, simple turing tests such as CAPTCHAs \citep{von2003captcha}, and more robust credentials \citep{adler2024personhood}. 
More generally, the goal of `know-your-customer' for users and granular access permissions (identity and access management, IAM) are commonplace on the internet.

Similarly, many websites seek to \emph{broadly limit access to bots on their services}, and may do so through the use of robots.txt bans. This is important since the widespread presence of bots or unauthenticated AI agents can lead to abuse and harm, but is often done at the `user-agent' level (for example, banning all `GPTBot' user agents~\citep{longpre2024consent}).

To \emph{track and verify the output of AI systems}, watermarking techniques \citep{liu_survey_2024, wang_data_2021} and content provenance measures \citep{c2pa_c2pa_2023} have emerged as potential solutions for determining the origin of AI-generated content. However, these approaches face reliability challenges \citep{saberi_robustness_2024} and are insufficient for establishing comprehensive accountability or safety when using AI agents. The inherent limitations of current verification methods highlight the need for more robust frameworks that can track not just content creation but also the broader implications of AI system deployment and interaction.

For \emph{managing access to sensitive AI capabilities} themselves, researchers have proposed `know-your-customer' schemes for compute providers \citep{egan_oversight_2023, obrien_deployment_2023}, while commercial platforms implement API tokens and access controls \citep{openai_chatgpt_2023}. These developments reflect a growing recognition that AI systems need robust mechanisms to prove their authenticity and permissions when accessing external services \citep{buterin_what_2023}, particularly as they become more integrated into critical infrastructure and decision-making processes.

To \emph{identify specific instances of AI agents}, recent work has proposed identifiers and verification approaches discussed above \citep{chan2024ids,chan2024visibility}. This is important and critical work, which we build upon to extend to \emph{authenticated delegation of AI agents} using existing authentication and permission protocols to enable AI agents to act on behalf of users in a controlled manner. In turn, these identifiers and delegation mechanisms can help create spaces that do not just gatekeep to human users but also enable AI agents to act on behalf of users with auditability and accountability.

\paragraph{Comparisons to Model Context Protocol and GPT Actions}
One example of an AI-centric protocol is the recent Model Context Protocol (MCP) \cite{anthropic2024mcp} from Anthropic, which enables secure, structured interactions between AI systems and external tools or data sources. MCP aims to enhance the contextual relevance of AI outputs by establishing a standardized framework for connecting models to resources to facilitate applications like retrieving live data, interacting with APIs, and executing tasks in real time.

While an extremely useful standard, it's limited in its full scope towards authorized delegation, enabling only system communication and optionally access controls rather than broader authentication and identity management.

Similarly, OpenAI's GPT Actions are integrations allowing GPT models to perform specific actions like booking a flight or retrieving data from APIs, which is a more constrained version of MCP and shares in its shortcomings. LangChain's Agent Protocol / LangGraph Platform extends this idea to enable multi-agent interoperability.

\paragraph{Documentation, safety, and governance of agentic AI systems}
Documenting AI systems and the data that create them has been a critical area of research and practice. Early frameworks established foundational approaches including datasheets \citep{gebru_datasheets_2021}, model cards \citep{mitchell_model_2019}, and data statements \citep{bender_data_2018}, with popular implementations emerging  \citep{paullada_data_2021}. Although each of these approaches has proven valuable, they face challenges in adequately addressing concerns around bias \citep{buolamwini_gender_2018}, privacy, and copyright. Recent work has highlighted the need for documentation of AI agents to understand their capabilities and limitations \citep{chan2024ids}, moving beyond static system descriptions to capture dynamic behaviors and interaction patterns. As AI systems become increasingly agentic, new frameworks are needed to document their evolving capabilities, decision-making processes, and potential risks \citep{bommasani_opportunities_2022}.

Recent work has explored runtimes for validating and reversing agent actions \citep{patil_goex_2024} and protocols for structured communication between language models \citep{marro_protocol_2024}. Researchers are also evaluating frontier models specifically for capabilities that could enable deceptive behavior \citep{phuong_evaluating_2024, fang_agents_2024}, while others advocate for tracking prior incidents \citep{wei_designing_2024} and establishing broader safeguards for AI agent interactions \citep{shavit2023practices}. Governance of AI agents is a rapidly evolving area of research and practice \citep{reuel2024open, kolt_governing_2024}, with increasing attention being paid to the development of frameworks that can ensure responsible deployment and operation of autonomous systems.

\paragraph{How authenticated delegation combines these solutions}
This work combines and extends these existing approaches---AI agent IDs and credentials, proof-of-personhood and identity verification for human users, and content provenance and watermarking methods---to form a cohesive framework. This approach inherits well-established practices for identity management while introducing explicit scoping and metadata for AI agents. This integration allows for granular, enforceable permission sets, clearer accountability chains, and richer context signals (like a model’s certifications or limitations) to be attached to each delegated action, with a more robustly verifiable construction than a simple agent ID system card. In effect, authenticated delegation complements existing standards and enhances their reliability by anchoring the actions of AI agents to verifiable human principals and recognized AI-specific credentials, creating a unified foundation for safe and accountable AI interactions. To this end, \autoref{sec:technical} introduces a concrete framework with additional security guarantees to package these elements together in a robustly verifiable way.

\section{Extending OpenID Connect for identifying and authenticating AI agents}\label{sec:technical}

To support the motivation of \autoref{sec:why}, this section proposes a concrete technical framework building on existing internet-scale authentication protocols to introduce mechanisms for delegating authority from users to AI agents and describes a token-based authentication framework that leverages OpenID Connect and OAuth 2.0. Our approach extends these battle-tested protocols to address the unique challenges of AI agent authentication while maintaining compatibility with existing internet infrastructure.

\subsection{OAuth2.0 and OpenID-Connect}
While new frameworks for AI system identification are emerging, there are valuable lessons to be learned from existing internet-scale authorization and authentication protocols. 
In particular, the {OAuth~2.0} protocol~\cite{RFC6749-Formatted} and its extensions provide battle-tested patterns for delegated authorization and identity verification that could inform the development of AI agent credential systems.

{OAuth~2.0} emerged from the need for users to provide authorization to one service to access resources located in another service, based on the RESTful paradigm~\cite{Fielding-2000-Thesis}.
A key requirement underlying {OAuth~2.0} is the ability for access to be continually granted even if later the user is absent (e.g., offline).
Existing user authentication protocols (e.g., MIT~Kerberos~\cite{RFC4120-Formatted}, CHAP~\cite{RFC1994-Formatted}) were developed primarily for the interaction between a human user utilizing a host computer connecting to the authentication server over the UDP layer.
The advent of the RESTful APIs meant that the parameters and flows had to be communicated over the HTTP layer, with the TLS providing the underlying message confidentiality layer.

A typical example of the scenario addressed by {OAuth~2.0} is the user who wishes to allow  an online calendaring service to read the user's itinerary from an airline service.
Here, the service that seeks access to the resource is referred to  as the {OAuth~2.0} {\em Client}.
On the other side, the service that is managing the resource is referred to as the {\em Resource Server} (RS). It is important to note that one of the assumptions underlying {OAuth~2.0} is the fact that Client and the RS can be operated by third-party entities.

The wide deployment and popularity of the {OAuth~2.0} protocol  enabled new features and extensions to be added. 
One successful extension---namely the {\em OpenID-Connect} protocol (OIDC)~\cite{OIDC1.0}---is the addition of flows dealing with the user authentication. The service dealing with authentication is referred to as the {\em OpenID Provider} (OP).
A key addition introduced by OpenID-Connect is the {\em ID-token}, which carries information about the human user that can be retrieved from the OP (i.e., by presenting ID-token). Here a merchant (as the Relying Party) would input the ID-token to the relevant token-validation endpoint at the OP in order to obtain more information about the user. We believe this capability may be extended to address the case of AI agents.

Another extension of the {OAuth~2.0} protocol that enables a user to manage multiple resources distributed across many Resource Servers is the {\em User-Managed Access} (UMA) protocol~\cite{UMACORE1.0}.
The UMA model may fit use-cases where the human user possesses multiple AI~Agents and where a single point of policy or rule configuration is desirable~\cite{Hardjono2019-IEEECommsMagazine}. Here, the AI~Agents can be viewed as distributed resource servers owned by the user.
Using the UMA~Authorization Server, the user can set policy at one location and have these policies automatically propagated to the multiplicity of AI~Agents.

\subsection{Delegation of authority from the user to the AI agent}

Given that the {OAuth~2.0} protocol is an authorization protocol, it is worthwhile considering reusing the {OAuth~2.0} patterns to establish a new mechanism for the human user to {\em delegate} certain tasks to the AI Agent. 
In other words, the human user is authorizing the AI Agent to carry out certain tasks that are limited in scope on behalf of the user.

In this new proposed extension, the human user must first perform authentication to the OpenID Provider (OP) to demonstrate their identity. The user then `registers' the AI Agent to the OP so that external entities who later seek to obtain further information about the AI Agent can do so to the OP. Registration could be done automatically in the background when an agent is created through a vendor (such as creating a new assistant instance with OpenAI).  

Existing {OAuth~2.0} client registration protocols can be customized to enable the user to register the AI Agent to the OpenID Provider and designate the AI Agent as a delegate or surrogate of the human user.

Next, the human user can issue a new {\em delegation token} that authorizes the AI~Agent to carry out tasks on behalf of the user.
Here, the term `authorize' is utilized to explicitly call out the fact that the AI~Agent is owned (driven) by a human delegator.

Both the user ID-token and the AI Agent delegation token can be referenced from within (or even copied into) a W3C Verified Credentials (VC) data structure~\cite{Sporny2022}.
This enables the AI Agent to wield the VC in its interactions with other entities (e.g.,  other services or other AI Agents), and have the benefit that both tokens would be verifiable at the standard OP.

It is worth noting that these delegation and authentication exchanges could alternatively be implemented using W3C VC issuance and delegation mechanisms. In such a scenario, a W3C VC could generate an OpenID-compatible credential, enabling seamless interfacing with OpenID systems. While this integration highlights the interoperability between W3C VC and OpenID ecosystems, further exploration and formalization of this process are left as future work and are beyond the scope of this paper.

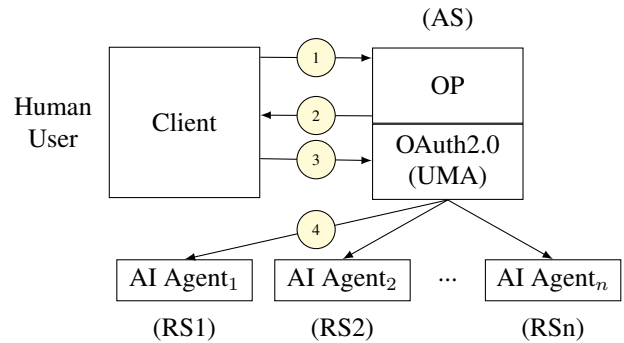
\begin{figure}[ht]
    \centering
    \begin{tikzpicture}[
        scale=0.7,
        box/.style={draw, minimum width=1.8cm, minimum height=0.5cm},
        user/.style={},
        circle_num/.style={circle, radius=0.2cm, draw, fill=yellow!20, font=\tiny},
        arrow/.style={->, >=latex}
    ]
    
    \node[box, minimum width=2cm, minimum height=2cm] (client) at (0,0) {Client};
    \node[left=0.1cm of client, align=center] {Human\\User};
    
    \node[box, minimum width=2cm, minimum height=1cm] (op) at (5,0.7) {OP};
    \node[box, minimum width=2cm, minimum height=1cm, align=center,  below=0 of op] (oauth){OAuth2.0\\(UMA)};
    
    \node[box] (a1) at (0,-3) {AI Agent$_1$};
    \node[box] (a2) at (3,-3) {AI Agent$_2$};
    \node at (5,-3) {...};
    \node[box] (an) at (7,-3) {AI Agent$_n$};
    
    \node[above=0.1cm of op] {(AS)};
    \node[below=0.1cm of a1] {(RS1)};
    \node[below=0.1cm of a2] {(RS2)};
    \node[below=0.1cm of an] {(RSn)};
    
    \draw[arrow] ($(client.north east)-(0,0.2)$) -- node[circle_num, midway] {1} ($(op.north west)-(0,0.2)$);
    \draw[arrow] ($(op.north west)-(0,1.3)$) -- node[circle_num, midway] {2} ($(client.north east)-(0,1.3)$);
    \draw[arrow] ($(client.south east)+(0,0.75)$) -- node[circle_num, midway] {3} (oauth);
    \draw[arrow] (oauth.south) -- node[circle_num, midway] {4} (a1.north);
    \draw[arrow] (oauth.south) -- (a2.north);
    \draw[arrow] (oauth.south) -- (an.north);
    
    \end{tikzpicture}
    \caption{Integration of OpenID Connect (OIDC) and User-Managed Access (UMA) protocols for establishing delegated authority from human users to AI Agents. The diagram illustrates the authentication flow where a human user first authenticates to an OpenID Provider (OP) (1 \& 2), registers their AI Agent (3), and issues a delegation token (4). This token empowers the AI Agent to perform authorized tasks on behalf of the user. The verification of both the user's ID token and the AI Agent's delegation token can be performed through the standard OpenID Provider, leveraging existing OAuth 2.0 patterns while incorporating new delegation mechanisms for AI Agent authorization.}
    \label{fig:OIDC-AI}
\end{figure}

\subsection{Token-based authentication framework}

Extending the existing OIDC framework, we can provide all relevant AI agent attributes and metadata of delegation in a set of identity-related tokens.

\begin{itemize}[topsep=1pt] 
\item	{\em User's ID-token}: This is the existing ID-token data structure that is issued/signed by the OpenID Provider (OP) service.
It is intended to represent information regarding the human user, and is no different to those used in everyday login experiences. 

\item	{\em Agent-ID token}: This carries the relevant information about that AI agent issued as an {OAuth2.0} Native Client (meaning the owner of the AI~Agent controls all keying material and secret parameters) and allows the corresponding service to verify any claims about the AI agent and its information. This token can include a range of additional information, from a unique identifier for the agent to a richer and more detailed agent ID token containing system documentation, capabilities or limitation metadata, relationship attributes to other AI systems, or other system characteristics. See \citet{chan2024ids} for further discussion of what an agent ID could entail.

\item {\em Delegation Token}: This newly introduced token explicitly authorizes an AI agent to act on the user’s behalf. 
The delegation token is issued and signed by the human delegator and carries references to (e.g., hash of) the corresponding user's ID token and the agent's Agent-ID token, allowing it to be verified by any service that trusts the OP.
Further, any relevant information about the nature of the delegation can be shared. For example, sharing the summarized goal of the agent and its scope limitations could assist a third party in guiding the AI agent to useful endpoints and interaction paradigms. The delegation token should specify validity conditions, such as expiration time or revocation endpoints, and be digitally signed by the user to prevent forgeries and ensure that the user knowingly granted the AI agent the listed privileges. 
In addition, the token may carry supplemental metadata—for example, logging or audit URLs—allowing service providers to record interactions, monitor delegated actions, and respond appropriately to anomalies.
By verifying that the delegation token references a valid user ID-token and a properly issued agent ID-token, remote services can confirm the authenticity and scope of the AI agent’s authority before granting access.
\end{itemize}

\subsection{Scope Limitations on Delegation}
The delegation framework enables human users to optionally define explicit boundaries for their AI Agents' actions by encoding scope limitations in the delegation token. However, given the flexible nature of agents and their diverse action space, scoping presents a unique and interesting challenge.

\subsection{Using verifiable credentials as an alternative} The W3C Verifiable Credentials (VC) standard \cite{Sporny2022} offers a versatile alternative—and sometimes complement—to existing OpenID Connect (OIDC) flows for conveying identity and delegation data. Under a VC-based approach, an issuer (such as an organization or individual) can sign a credential that attests to various claims about a subject, which might be a user, an AI agent, or any other entity needing verifiable, tamper-evident attributes. Because VCs are not bound to a particular transport protocol, they can be presented and verified in a decentralized or peer-to-peer manner without always relying on a single identity provider. This contrasts with OIDC, which generally depends on a central OpenID Provider (OP) to mint and validate tokens.

A key benefit of VCs is their privacy-enhancing potential. Rather than disclosing all attributes or relying on a single identity provider, users, and AI agents can share only the subset of claims strictly necessary for a given interaction. This ``selective disclosure'' capability can mitigate concerns around centralized logging or cross-platform correlation inherent in OIDC-based architectures, especially when interactions span multiple domains or organizations.

Nonetheless, replacing OIDC entirely with a purely VC-based model does come with trade-offs. OIDC already enjoys a robust ecosystem of libraries and deployments that provide well-tested support for issues like token refresh, revocation, and audience restriction. VCs, while powerful, require additional work to replicate these flows at scale—particularly if each verification call demands a new signature check or interaction with a blockchain or distributed ledger. In many enterprise environments, stakeholders may prefer to incorporate VCs into existing SSO or multi-factor authentication frameworks, rather than adopt a fully decentralized identity infrastructure upfront.

In practice, hybrid solutions often prove the most pragmatic. A user or AI agent could store and manage VCs encoding rich attributes or regulatory endorsements, while still leveraging OIDC tokens to bootstrap compatibility with existing authentication or authorization endpoints. For instance, an Agent-ID token could embed a VC carrying detailed metadata on its behavioral, property, context, and relationship attributes. Service providers integrating with OIDC get the familiar token-based handshake, while still retaining the option to parse the embedded VC for an additional layer of trust and context. Examples such as OID4VC support this~\cite{yasuda2022openid}.

\section{Defining scope and permissions for AI agents}\label{sec:scoping}

Authenticated delegation is inherently tied to robust scoping mechanisms, as users must be able to specify their permissions and instructions in a clear and unambiguous manner. This comes in direct conflict with the extremely large possible action space AI agents can perform. 

While much work in reliability and alignment focuses on ensuring that AI agents follow instructions correctly, the risks of misinstruction, prompt injection attacks, and reduced security auditability make pure natural language prompts an incomplete scoping, permission, and security tool. \textbf{This section addresses how AI agent infrastructure can bridge the gap between these natural language instructions and robust concrete access control mechanisms} by proposing converting flexible natural language scoping instructions into machine-readable, version-controllable, and auditable structured permission languages that can be leveraged for use in human-in-the-loop settings.

We distinguish between \textbf{task scoping} and \textbf{resource scoping}:
\begin{itemize}[noitemsep, topsep=0pt] 
    \item Task scoping involves specifying which actions or workflows an agent is authorized to perform on behalf of the user. These actions may range from high-level tasks (e.g., ``draft a financial report”) to more granular actions (e.g., ``create a new database entry”);
    \item Resource scoping involves specifying which resources (information, APIs, tools, etc.) an agent can use or modify.
\end{itemize}
While conceptually distinct, task scoping and resource scoping are closely connected. Limiting which tasks can be performed also means that a (well-designed) agent will not access unnecessary resources; similarly, restricting access to specific resources also constrains what tasks are feasible in the first place.

This section addresses how access control mechanisms can be integrated with complex AI agents and natural language workflows. It outlines the critical nature of structured permissions, how they can provide a robust and generalizable foundation for agent scoping, and how natural language and human oversight can be flexible interfaces for these access controls. 

\subsection{Structured permission languages}

A large class of scoping mechanisms relies on structured, machine-readable policy specifications. These specifications unambiguously define which entities have which authorizations, under which conditions, and with what privileges. Several well-known languages and frameworks exist for encoding permissions, such as XACML (eXtensible Access Control Markup Language), which uses XML to encode and evaluate access control policies \citep{xacml}, and ODRL (Open Digital Rights Language), designed for expressing usage permissions over digital content \citep{odrl}.
Other languages include OBAC \citep{obac}, ROWLBAC \citep{rowlbac}, KaOS \citep{kaos} and Multi-OrBAC \citet{multiOrbac}, which rely on ontologies (typically described using OWL) to model resources, subjects, and authorizations. In web-based contexts, this can often be as simple as whitelisting or blacklisting URLs and subdomains that an agent can access.

These structured languages are machine-readable and can thus be enforced reliably by traditional (non-AI) systems.
From a practical perspective, they are well-suited for resource scoping, since resources are typically discrete and can be classified, enumerated, and grouped into security domains. For instance, when a policy states that a certain directory is read-only for a particular agent, enforcing compliance is straightforward and can be implemented at the system level.

However, they have three main drawbacks. 
First, while these frameworks are suitable for enumerating resources, they are less flexible for task scoping, especially when tasks are open-ended or cannot be easily described as a set of operations. 
Second, policy definitions can become lengthy and complex, especially in environments with a large number of resources and tasks, or in web contexts where the number of possible web interactions is enormous.  
Third, they are often environment-specific and require updating for different digital systems with which an agent interacts. 

Despite these drawbacks, structured permission languages remain a cornerstone of access control because they provide a precise, easily auditable basis for resource scoping. An alternative approach involves using \emph{schema validation} to constrain how agents interact with the environment, discussed in \autoref{sec:schemavalidation}

\subsection{Authentication flows}

Another dimension of controlling agent behavior is the \textbf{authentication flow} (i.e., deciding when to prompt a user or another authority for confirmation before the agent proceeds with an action). Rather than frontloading all access decisions into a single policy definition, an authentication flow can dynamically request user approval for borderline or high-risk operations.

The main advantage of this approach is that users do not need to define every edge case in a static policy. Additionally, authentication flows can be combined with other scoping mechanisms: for example, a policy can state that any resource that is neither explicitly approved nor explicitly forbidden requires human approval.

On the other hand, frequent authorization prompts can negatively affect the user experience, leading to ``prompt fatigue''~\cite{securitypromptfatigue}, where the user simply grants permissions without a proper review. Moreover, determining when a request requires explicit authorization can be non-trivial, and misclassifications can lead to either excessive prompting or critical operations slipping through unnoticed.

In practice, a well-designed system can combine robust, structured policy definitions (for common scenarios) with dynamic authentication flows for rare or particularly sensitive actions. This approach allows users to offload the majority of routine checks to automated policies while still preserving the ability to escalate novel or ambiguous requests for user confirmation.

\subsection{Natural Language Mechanisms}

Alongside fine-tuning, prompting has often been employed to steer the behavior of a model towards safety \citep{zheng2024prompt}. A reasonable extension of this approach would be to train (or prompt) the LLM to interpret permissions described in plain language.
For instance, a user might say, ``You are allowed to generate summaries of public documents, but you must not reveal any confidential metrics.'' Such instructions can, in principle, be parsed and acted upon by an LLM-based system.

The main strength of this paradigm is its user-friendliness. Non-technical users may find expressing policies in natural language much easier than writing formal rules. Moreover, natural language can capture nuanced or context-dependent instructions that are difficult to encode in structured languages. This makes them ideal for both task and resource scoping.
Finally, natural language can be used to enforce policies on actions that require reading or using natural language, such as interactions with other LLM-based agents.

However, natural language often lacks the precision needed for reliable policy enforcement. For instance, terms like ``sensitive data'' or ``private emails'' may be interpreted differently depending on context. This problem is particularly relevant in the case of conflict between different policies, where ambiguous and context-dependent instructions may yield different interpretations.
Relying solely on an LLM to interpret and enforce ambiguous natural language instructions can be risky in security-sensitive contexts.

In short, while natural language instructions can serve as a convenient mechanism (especially for task scoping, where other mechanisms are less suitable), they are \textbf{not} reliable enough to be used as standalone policy mechanisms.

\subsection{Combining structured permissions, natural language, and user oversight}

\paragraph{Resource scoping as a foundation.} We argue that the most broadly applicable strategy is to enforce \emph{resource scoping with structured permissions}. The brittleness of natural language mechanisms makes them unsuitable for production-level usage of AI agents, especially when security or compliance is a concern. In contrast, structured permissions are unambiguous and deterministic, providing verifiable guarantees against unauthorized access. Focusing on resource scoping also significantly reduces the overhead of specifying every authorized task in detail. To an extent, agents could attempt to represent task-scoping instructions in the form of resource scoping, using domain knowledge of the contexts in which they operate.  
Since resources are generally discrete and can be classified, enumerated, and grouped into domains, controlling resource access implicitly prevents many potential tasks that would require out-of-scope resources.
Additionally, structured resource scoping has several advantages:
\begin{itemize}[noitemsep, topsep=0pt] 
    \item It does not depend on how a user delegates tasks---be it via a script, an AI agent, or a more traditional workflow;
    \item It is more compatible with existing non-AI access control systems, which focus on machine-readable permissions for resources (e.g., databases or URLs);
    \item It is suitable for structured logging and version control, which simplifies auditing and compliance reporting.
\end{itemize}

Though users may supplement resource scoping task constraints written in natural language, the core resource-based policies provide a safety net that is largely immune to ambiguities in language or model vulnerabilities. Even if an LLM or another AI agent is tricked or misaligned, its ability to execute harmful actions is constrained by the underlying resource permissions.

\paragraph{Connecting to natural language.} While robust and auditable, structured resource scoping alone lacks ease of use and flexibility. To address this, the instructions for the LLM (or a separate scoping prompt) can flexibly express the scoping limitations that should be applied. These natural language scopes can be converted to a structured scoping format by the agent or an AI system in the corresponding environment (which has more detailed knowledge of the relevant resource profiles). Examples of conversion between natural language and structured permissions include \citet{subramaniam2024intent}, which generates PostgreSQL restrictions, and \citet{jayasundara2024ragent}, which uses retrieval to generate custom JSON policies.

A similar process could also be performed for different environments and digital services an agent interacts with, allowing a flexible set of permission instructions to be applied across a wide range of services and contexts (which is important given the broad action space of AI agents).

\paragraph{Bringing a human in the loop.} The key final step is validating these structured access controls via the human delegator.  
Authorization workflows present an opportunity for users to briefly review and approve structured access control limitations for different systems. For instance, in \citet{wright2024here} LLM agents agree on structured information (in this case, meeting dates) which are then confirmed by human users.

\paragraph{Combining into a hybrid implementation.} 
Bringing these elements together into an implementation is relatively straightforward. An LLM assists in converting high-level, natural language resource constraints into formal, structured rules that users can subsequently review and approve.
\textbf{For example:}
\begin{enumerate}[noitemsep, topsep=0pt]
    \item A user writes: ``Allow the agent to read and write to the directories about `projectAlpha`, but do not grant it access to the folders with financial folders;”
    \item The LLM translates this requirement into a policy definition, either in a universal permission language (e.g., XACML) or in the specific permission language used by the resource (e.g., SQL access policies for databases). In this specific case, the LLM enumerates ``projectAlpha” resources while explicitly denying access to ``financials2023;''
    \item The user reviews, corrects if necessary, and finalizes the policy.
\end{enumerate}
While many specific details of such a workflow need to be address such as intermediate validation checks and the evaluation of robustness of LLM translation into structured languages, we leave these specifics to future work.

Ultimately, focusing on structured, unambiguous resource constraints is the most reliable way to ensure that an AI agent remains within authorized bounds in a given environment. While there is still room for higher-level (often natural language) task constraints, one should treat these as guidance towards the primary enforcement mechanism. Indeed, while natural language can adequately address the extremely large possible space of agent actions, its transformation into access controls grounds the limitations on agent actions into finite auditable controls. 
Structured resource scoping reduces the reliance on model alignment alone, decreases the risk of adversarial prompt injections, and simplifies the integration with well-established security mechanisms. 
Combining this approach with well-designed authentication flows and helping the user interpret the generated policies can reduce the chances of human errors, enhance accountability, and improve the robustness of authenticated delegation. 

\subsubsection{Inter-agent scoping.}
Extending beyond the user-agent-service model, this approach can apply to multi-agent settings where agents want to propagate their limitations onto other agents performing actions on their behalf.
Suppose that the user specifies the authorizations of an agent Alice. When Alice interacts with another agent, Bob, in natural language to perform a task, Bob can parse Alice's scoping instructions and interpret them in its own environment. 
By doing so, Bob can confirm that its assigned operations remain within the original scope, and provide an auditable receipt of the actions taken and the resources accessed. This is particularly useful in scenarios where inter-agent communication spans different organizations, each with separate policies and resource constraints.

For a concrete example, suppose that Alice is a project management agent and Bob is an accounting agent. The user describes in plain English a financial data request to Alice; Alice thus sends the forwarded request and a description of the authorizations to Bob. Bob replies with a structured interpretation of the authorizations (e.g., ``Read-only access to `transactions2025' dataset, columns: total amount, vendor name''), which is logged and approved by either the user or Alice.

Such a workflow ensures that even if the agents communicate in flexible natural language, their underlying scoping and record-keeping remain anchored in auditable, deterministic policy. As a result, the risk of unauthorized data sharing or unbounded agent behavior is greatly reduced, and each agent’s capacity to ``inherit'' restricted credentials from the delegator is tightly controlled.

\section{Discussion}

\subsection{Problems with an OpenID Connect approach} \label{sec:shortcomings}
While the OpenID Connect (OIDC) and OAuth 2.0-based framework proposed here provide robust and battle-tested mechanisms for authentication and delegation, it comes with trade-offs and may be more complex than alternatives with different trade-offs in privacy, security, and auditability.

\paragraph{Overhead from multiple sign-in flows.} A significant drawback of the OpenID Connect approach is the potential overhead introduced by multiple sign-in flows required to authorize AI agents across individual service providers. This can be likened to the experience of setting up a new email client, where users must repeatedly log in to authorize access to various services. While such authorization flows enhance security by ensuring each provider independently verifies the AI agent's delegation credentials, they impose a usability cost by slowing down access to secure systems. In theory, it is possible to bypass this burden by presenting delegation tokens directly without performing the full OIDC authentication flow; however, this shortcut sacrifices key security guarantees, particularly those related to token freshness and verification.

\paragraph{Increased reliance on OpenID Providers and privacy risks.} The reliance on OpenID Providers (e.g., Google, Facebook, or equivalent entities) introduces systemic privacy concerns. Since OIDC providers mediate all authentication flows, they gain the ability to track and correlate individual AI agent interactions across various services. This can include collecting statistical usage analytics or requiring relying parties to share logs, which facilitates extensive behavioral profiling. Such centralized visibility undermines user privacy and creates a potential single point of surveillance. Addressing these risks necessitates strong privacy mitigations, such as pairwise pseudonymous identifiers or the minimization of log-sharing requirements, but these mechanisms add further complexity to the system.

\paragraph{Comparative complexity relative to W3C Verifiable Credentials.} While the paper highlights the ability to embed W3C Verifiable Credentials (VC) within the OIDC framework, the full OIDC authorization flow may still be unnecessarily heavy compared to native W3C VC-based delegation and authentication processes. W3C VC issuance, authentication, and delegation mechanisms could directly fulfill the same requirements for AI agent identity verification without incurring the additional overhead of repeated authorization flows and central provider mediation. Additionally, W3C VC-based approaches are inherently more privacy-preserving, as they do not rely on a single provider to mediate trust or track credential usage. A streamlined VC-based process could generate OIDC-compatible credentials when required, enabling interoperability while preserving simplicity and privacy. 
Similarly, other proposed alternatives to OAuth 2.0 specifications could be drop-in solutions here to address design trade-offs, such as the Grant Negotiation and Authorization Protocol (GNAP) \cite{rfc9635}.
Further exploration of these alternative approaches remains essential to determine their feasibility as lightweight solutions for AI agent delegation. 

Taken together, these limitations highlight key trade-offs between security, usability, and privacy in the OIDC-based framework. While the proposed approach remains an incremental and interoperable path forward, addressing these challenges will be critical to ensuring a robust and practical system for AI agent authentication and delegation.

\subsection{Limitations of natural language scoping}

Although translating natural language scoping instructions into structured permission languages enables a more flexible interface, it also creates several key challenges.

\paragraph{Evaluating reliability and correctness.} One of the foremost difficulties is ensuring that the translation from a user’s natural language specification to a machine-readable policy is accurate. Natural language instructions often contain context-dependent or ambiguous terms, making them inherently prone to misinterpretation by an AI system. Although a human-in-the-loop approach can mitigate these risks through policy review, such human verification is not infallible; users may inadvertently miss subtle translation errors.
Moreover, as the complexity of a permission specification grows, verifying the alignment between the original natural language instruction and the generated structured policy becomes more difficult, both technically (due to large policy definitions) and cognitively (due to the burden on human reviewers).

\paragraph{New threat vectors for LLM attacks.} Exploiting weaknesses in language-based interfaces can expose novel threats that do not exist under purely static access control.
Prompt injection and jailbreak attacks can coerce a large language model into generating or accepting policies that exceed the original user’s intent, thereby gaining unauthorized privileges. While separating resource or task-scoping instructions from normal chat sessions or interactions reduces the likelihood of an attack, it still presents a new differentiated attack surface that needs to be guarded.

\paragraph{Contextual drift.} As policies evolve or the task context changes over time, prior natural language instructions risk becoming outdated or misaligned with newly introduced resources. Maintaining consistency across multiple revisions of instructions is nontrivial.

\paragraph{Partial reliance on third parties to enforce restrictions.} In some contexts, the access control rules are applied to an external environment or agent that is being interacted with. To maintain security over the application of these access controls, it may be necessary for the corresponding party to enforce the rules beyond trusting the native agent to follow them. In such instances, the reliability of the third-party becomes a critical point of failure. 

\subsection{Can model vendors provide this?} \label{sec:bg-AI-labs}
Model vendors (e.g., OpenAI, Anthropic, Google) can provide tooling to share which user is being represented when an AI system accesses a digital service and what the intended scope or permissions are. This is encouraged. 
However, current approaches to sharing such information are insufficient from a security and verifiability perspective, such as including the information in the user-agent string of the AI system, or writing the information into API calls made by the AI system. Instead, these services could act as an OpenID Provider (or partner with one) for the AI system without any change to the user experience, or if they prefer a different instantiation of the authenticated delegation framework, they could provide W3C verifiable credentials paired with robust, unique IDs for AI agents and users.

Implementing authenticated delegation is also feasible when AI systems and agents are self-hosted or deployed on custom infrastructure. This includes leveraging internal identity management infrastructure for human users and incorporating custom permission controls. Such systems can operate internally within an organization to ensure AI system usage aligns with identity and access management (IAM) policies and delegation frameworks across various technology stacks and modalities.

\subsection{How this interacts with robots.txt}
Robots.txt has, without legal heft, underpinned the modern web for decades. It relies upon a simple set of directives, where a user-agent is given rules for a subroute. Just as the recent proliferation of scraping has led to rapid uptake of new user-agent rules \citet{longpre2024consent}, new directives could easily be rolled out across the web with the right incentives.

This system still has a place in a web full of AI agents. While websites may wish to block scraping, they may also wish to guide agents to the correct subroutes where they could share credentials and interact. For example, a website may wish to block scraping, allow human users to interact, and send AI agents directly to an API natural language interface designed for AI systems.

To guide agents to the correct subroutes where they could share credentials and interact, we can define a new user agent, \texttt{AgentBot}, and force it into a specific interaction route (e.g., \texttt{/AgentInterface/}). Since \texttt{robots.txt} is a guide, not a rule, this route can go on to provide richer details of what services can be accessed and what sitemaps exist. Such a \texttt{robots.txt} need only be an initial guide to agents.

\subsection{Legal grounding for authenticated delegation} \label{sec:legal}
The law of agency addresses circumstances in which one party, the principal, authorizes another party, the (human) agent, to act on their behalf~\cite{blackslawdict}. 
At its core, agency law determines when a principal may be held liable for the acts of their agent, ensuring that third parties are not unfairly disadvantaged by having to ascertain who holds ultimate responsibility. 

A key result of agency law is to instill trust and confidence in market transactions: by providing clear rules about liability and authority, agency law reduces uncertainty and contributes to more efficient market operations~\cite{posner_econ_analysis_law,williamson_markets_hierarchies,casadesus2005trust}. 

One central concept in agency law is that of ``apparent authority,'' extensively discussed in the Restatement (Third) of Agency~\cite{restatement_third_of_agency}. Under this doctrine, a principal can be held responsible for acts that a reasonable third party perceives the agent to be authorized to perform, even if the principal never granted that authority explicitly. This principle also helps maintain market stability: third parties need not investigate every aspect of an agent’s credentials or verify each claim of authority before proceeding with a transaction, as long as the agent appears to be acting on behalf of the principal in a reasonable manner. 

It remains uncertain how established agency doctrines will adapt to AI agents that can learn, self-modify, or operate autonomously~\cite{balkin2015path, adler2024personhood}. Traditional notions of intent, consent, and observable authority are difficult to apply to current autonomous systems. In response to these uncertainties, the authenticated delegation framework offers a model in which each delegation of authority is verifiable. Rather than relying on appearances, this framework enables third parties to automatically confirm that an AI agent is indeed authorized to act on behalf of a principal. In doing so, it reduces the need to rely on apparent authority doctrines and diminishes the risk of mis-attribution of actions. 

A recent controversy involving Air Canada illustrates how these principles might play out in practice~\cite{BCCRT2024}. In this instance, the airline argued that it could not be held liable for information provided by its online chatbot. Implicitly, this suggests treating the chatbot as if it were separate from the airline—akin to an independent entity. Yet, in the judge's view, the chatbot exists as part of Air Canada’s digital infrastructure and so the company was responsible for the information it provided. Under conventional principles of law and equity, the chatbot’s outputs, even if generated autonomously, form part of the information the airline holds out to the public. The airline’s attempt to evade responsibility runs counter to the principle that a firm must stand behind the representations it makes, whether through humans or machines. This case underscores that companies may be liable for the actions of their AI agents, a view also held by many scholars~\cite{adler2024personhood}. From a broader perspective, this case also highlights the growing need for robust technological and legal mechanisms---like the authenticated delegation framework---that can delineate responsibility and authority in AI-mediated interactions, ultimately protecting consumer trust and market stability.

Beyond agency law, existing legal frameworks for electronic transactions, like the Uniform Electronic Transactions Act (UETA), provide some guidance. The UETA is a uniform law adopted by 49 U.S. states to help accommodate the realities of e-commerce by recognizing that electronic communications and automated processes can play substantive roles in forming and executing agreements~\cite{transactions_on_agents,ueta_official_text}. Under UETA, parties are encouraged to adopt agreed-upon security procedures and error-detection protocols to ensure that the electronic records genuinely reflect the intended agreements. If one party fails to follow these procedures and an error that would have been detected goes unnoticed, the other party may be permitted to avoid the consequences of that error. Similarly, if an individual errs while interacting with an electronic agent and the system offers no reasonable correction mechanism, UETA contemplates relief for that individual under defined conditions.

These provisions reflect an understanding that trust in digital commerce requires more than just a willingness to be bound by electronic contracts; it also demands reliable methods for verifying authority, correcting mistakes, and ensuring that automated processes faithfully implement the intended instructions of the principal. The authenticated delegation framework aligns well with these goals. By integrating a verifiable chain of authority into interactions with AI agents, it provides the digital equivalent of an agreed-upon security procedure. In doing so, it can reduce misunderstandings and disputes about whether an AI-driven process was acting within the scope of its authority. 

A critical element of both trust and accountability in AI-augmented systems lies in maintaining meaningful human oversight, often termed the ``human-in-the-loop'' requirement. The EU AI Act, for example, emphasizes the importance of maintaining human involvement in high-risk AI decisions to ensure ethical, transparent, and accountable outcomes~\cite{eu_ai_act}. The authenticated delegation framework supports this principle by making the human role in agent workflows explicit. Rather than delegating authority to an AI system behind opaque layers of code, third parties can firmly establish when, how, and under what conditions the AI is authorized to act. This allows humans to step in to verify decisions, correct errors, and ensure that automated actions remain aligned with overarching legal and ethical standards.

The interplay between technology and law in the context of AI-driven agents is complex and evolving. Strengthening the legal underpinnings, adopting frameworks for authenticated delegation, and integrating human oversight at critical junctures are all steps toward ensuring that emerging AI systems not only enhance market efficiency but also maintain core values of trust, fairness, and accountability. Further empirical and doctrinal analysis could deepen this conversation, drawing on works that examine the real-world implementation of human-in-the-loop mechanisms~\cite{mosqueira2023human}.

\section{Conclusion}
This paper presents a practical framework for authenticated delegation to AI agents, addressing urgent challenges around authorization, accountability, identity verification, and access control management in digital spaces. By extending existing OAuth 2.0 and OpenID Connect protocols with AI-specific credentials and delegation mechanisms, our approach enables secure delegation of authority from users to AI agents while maintaining clear chains of accountability. The proposed token-based framework - comprising user ID tokens, agent-ID tokens, and delegation tokens - provides a robust foundation for verifying agent identities, controlling permissions, and maintaining audit trails, while supporting granular and robust scope limitations generated in response to natural language instructions. 
Our key contribution is demonstrating how established internet-scale authentication (e.g., OpenID Connect and W3C VCs) and access management protocols (e.g., XACML) can be adapted to address the unique challenges of AI agent delegation while preserving compatibility with current systems, as illustrated through real-world use cases in areas like automated negotiations and web service interactions. As AI agents become more prevalent in digital spaces, frameworks like this will be essential for ensuring they operate within appropriate bounds while remaining accountable to their human principals. Looking ahead, key research directions include developing standardized scope definitions for common AI agent tasks, exploring privacy-preserving delegation mechanisms, and creating tools to help service providers implement and manage agent authentication policies, ultimately working toward ensuring AI systems can be safely and productively integrated into existing digital infrastructure.

\section{Acknowledgements}
We thank Kim Hamilton Duffy, Ankur Banerjee, Steve McCown, Shrey Jain, and Raina Wu for their feedback and support on this work.

\bibliography{aistuff,hardjonoshort,legalstuff,scoping}
\bibliographystyle{icml2025}

\newpage
\appendix
\onecolumn

\section{Technical Details}

\subsection{Federated OpenID Providers for Agent Mutual Authentication}







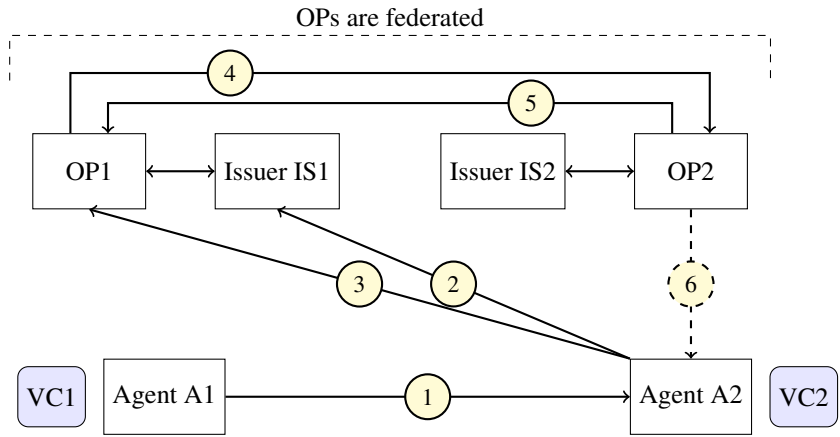
\begin{figure}[h!]
	\centering
	\begin{tikzpicture}[
		box/.style={draw, minimum width=1.5cm, minimum height=1cm},
		agent/.style={box},
		issuer/.style={box},
		op/.style={box},
		circle_num/.style={circle, draw, fill=yellow!20, font=\small},
		vc/.style={draw, rounded corners, fill=blue!10, minimum width=0.8cm, minimum height=0.8cm},
		arrow/.style={->, thick}
	]

	\node[op] (op1) at (2,3) {OP1};
	\node[issuer] (is1) at ($(op1)+(2.5,0)$) {Issuer IS1};  
	\node[op] (op2) at ($(op1)+(8,0)$) {OP2};  
	\node[issuer] (is2) at ($(op2)+(-2.5,0)$) {Issuer IS2};  
	
	\node[agent] (a1) at ($(op1)+(1,-3)$) {Agent A1};
	\node[agent] (a2) at ($(op2)+(0,-3)$) {Agent A2};
	
	\node[vc] (vc1) at ($(a1)+(-1.5,0)$) {VC1};
	\node[vc] (vc2) at ($(a2)+(1.5,0)$) {VC2};
	
	\draw[arrow] (a1) -- node[circle_num, midway] {1} (a2); 
	\draw[arrow] (a2.north west) -- node[circle_num, midway] {2} (is1.south);  
	\draw[arrow] (a2.north west) -- node[circle_num, midway] {3} (op1.south); 
	\draw[arrow, dashed] (op2) -- node[circle_num, midway] {6} (a2); 
	
	\draw[arrow] ($(op1.north)+(-0.25,0)$) -- ++(0,0.8) 
	          -- ++($(op2)-(op1)+(0.5,0)$) node[circle_num, pos=0.25] {4} -- ($(op2.north)+(0.25,0)$); 
	\draw[arrow] ($(op2.north)+(-0.25,0)$) -- ++(0,0.4) 
	          -- ++($(op1)-(op2)+(0.5,0)$) node[circle_num, pos=0.25] {5} -- ($(op1.north)+(0.25,0)$); 

	\draw[arrow, <->] (op1) -- (is1);
	\draw[arrow, <->] (op2) -- (is2);

	\draw[dashed] ($(op1.north west)+(-0.3,0.7)$) -- ++(0,0.6) 
		-- ++($(op2.north east)-(op1.north west)+(0.6,0)$) node[midway, above] {OPs are federated} -- ($(op2.north east)+(0.3,0.7)$);	


	
	\end{tikzpicture}
	\caption{Authentication flow between federated AI Agents demonstrating cross-domain credential verification. Agent A1 presents its Verifiable Credential (VC1) to Agent A2 (step 1), followed by claim validation through IS1 (step 2) and OP1 authentication (step 3). The federation network enables OP1 to validate A2's credentials with OP2 (steps 4-5), concluding with A2's credential presentation (step 6). The architecture supports secure authentication between agents operating under different OpenID Providers.}
	\label{fig:OP-federation}
\end{figure}

One of the key goals of AI~Agents is the ability for an agent to interact with existing web services as well as other AI~Agents (AI~Systems). To enable secure interactions, AI~Agents must perform mutual authentication and verify each others' credentials, including Agent-ID tokens and delegation tokens. \autoref{fig:OP-federation} illustrates this authentication process in a federated environment.

The authentication flow begins when agent A1 presents its Verifiable Credential to agent A2. The VC contains claims that must be validated through the respective OpenID Provider, including the user's ID-token and the Agent-ID token. Since the APIs at OP1 are protected, A2 must authenticate itself using its own Agent-ID token previously issued by OP2 in its home domain. This cross-domain verification is achieved through federation, where OP1 validates A2's credentials by communicating with OP2. While the figure demonstrates authentication from A1's perspective, the process is mutual, ensuring both agents can verify each other's delegated authorities and credentials through their respective OpenID Providers.

\subsection{Identification of AI~Systems and AI~Agents}\label{sec:global-identifiers}

One of the challenges facing the deployment of AI technologies
is the need to establish identification mechanisms for instances of AI systems, including AI Agents \citep{chan2024ids}.
%
Here it is useful to distinguish two basic types of identifiers:

\begin{itemize}
\item	{\em Local identifiers}:
A local identifier is a unique string (e.g. UUIDv2)
that can be used to distinguish an instance of an AI system from another within a given domain. 
This means that other systems and entities in the domain are able to pinpoint 
each AI system using that local identifier. 
A local identifier may be meaningless outside the domain, 
and thus require a mapping to a global identifier.

\item	{\em Global Identifiers}:
A global identifier enables an AI system to be referred to (or referenced to) from anywhere in the Internet. 
This enables agents to interact with other AI systems and other AI agents across different geographies. 

From a scalability perspective, it is useful to be able to map from the global identifier of an AI agent 
to its local identifier to enable other systems within its home domain to
provide support for that AI agent, 
such as a local authentication by the OP in that home domain 
that attest to the true existence of the agent within the domain. 

A global identifier belonging to an AI system or agent can be 
incorporated within a {\em decentralized identifier} (DID) structures~\cite{W3C-DID-2021}
that then enables useful interactions with DLT based services that function based on the DID.
\end{itemize}

Due to the prevalence of {OAuth~2.0} and OIDC deployments today,
it is useful to reuse some of the existing identifier structures 
already utilized in these deployments.
If we view an AI~Agent as being a client (native or hosted service) within {OAuth~2.0}
then we could reuse the two important parameters 
used by an {OAuth~2.0} client to interact with the authorization server (or the OP). 
These parameters are the {\tt client\_id}
and the {\tt client\_secret} (see section 2.3.1 of~\cite{RFC6749-Formatted}). 
The client-ID and the client-secret parameters in {OAuth~2.0}
is used by the authorization server (the OP) to recognize a client that had been previously 
registered to the OP using specific client registration protocols~\cite{RFC7591-Formatted}. 
In the current context of identifying AI systems and AI agents, 
the client-ID could be considered a local identifier that is meaningful only 
in the domain serviced by the specific OP (i.e. where the client has been registered). 
However, the client-ID could be the basis for the OP to issue a {\em delegation token}
that signifies the user delegating authorization to their AI~Agent to carry out certain tasks,
defined by {\em action scopes} within the delegation token.

\subsection{ID token threat model}

Our proposal is meant to be secure against several different security threats, ranging from the authenticity of the issued tokens to the nature of the delegation.

With respect to ID tokens, \citet{chan2024ids} identifies three fundamental threats that an AI ID system must defend against. The first threat is tampering, where an attacker modifies the ID while it is being transmitted between the author and the receiving party, potentially altering crucial system information or attributes. The second threat is ID spoofing, where an attacker creates a fraudulent ID and falsely claims it originated from a legitimate author (such as a major AI company), which could enable malicious systems to masquerade as trusted ones. The third threat is instance spoofing, where an attacker takes a legitimate ID and attempts to use it with their own unauthorized AI instances, essentially hijacking the reputation or privileges associated with the original system. To counter these threats, the authors propose that IDs must implement digital signatures that cover both the ID itself and the system's outputs, similar to how HTTPS certificates work for websites. However, they note an important limitation: since the signature must cover both ID and output, any modification to the output (even benign ones) would invalidate the ID, creating a challenging trade-off between security and usability. These threats to robust AI system identification extend naturally to the task of authenticated delegation for AI agents, which requires robustness for both AI system verification, human delegate verification, and verification of valid delegation.

OpenID Connect could help prevent several additional threats beyond these robust IDs. Through its built-in mechanisms, OIDC could prevent identity correlation attacks by using pairwise pseudonymous identifiers to ensure AI instances appear different to different services, thwarting attempts to track instance behavior across platforms. Its session management capabilities could prevent session hijacking attempts against active AI instances, while its dynamic client registration could prevent impersonation through unauthorized endpoints. Most significantly, OIDC's scoping and audience restriction mechanisms could prevent authorization scope abuse and cross-instance privilege escalation, ensuring AI instances cannot exceed their intended permissions or use tokens meant for other instances. The protocol's discovery mechanisms could also prevent identity provider spoofing, adding another layer of security to the ID ecosystem.







\section{Alternatives in Access Control Management}

\subsection{Schema Validation As Scoping} \label{sec:schemavalidation}

An alternative approach to structured permission languages involves using \emph{schema validation} to constrain how agents interact with the environment. In this approach, an AI agent’s possible outputs or queries must conform to a predefined schema \citep{allemang2024increasing}. For example, if the agent can only communicate using RDF tuples, the system can enforce rules on the permissible classes, properties, or relationships that the agent can generate.


In practice, schema validation can be particularly powerful in scenarios where the system is designed around standardized data formats (e.g., JSON, XML, RDF). By restricting the agent to these formats and validating every output (e.g. using JSON-Schema \citep{jsonSchema} or SHACL \citep{shacl}), schema validation indirectly controls which actions are feasible. For instance, if an agent is only allowed to generate RDF triples with certain predicates (e.g., ``hasTitle” or ``hasSummary”) and certain classes (e.g., ``Document”), it cannot arbitrarily mutate data outside of that schema domain.

Like structured permission languages, non-AI systems can quickly and deterministically verify if an agent's output complies with a given schema. Moreover, since the output of the agent is already structured, schema validation may be simpler compared to parsing unstructured text. Standard outputs also simplify logging, as every action can be captured and audited with structured queries.

On the other hand, a rigid schema can reduce flexibility, especially in the context of task scoping. Tasks that require nuanced or creative outputs can be difficult to capture in a schema-based approach without introducing significant complexity (especially if such tasks evolve over time). Moreover, designing a robust schema that is both expressive and safe requires considerable effort, and the agent must be trained or prompted to work exclusively within that schema.

Nevertheless, schema validation can be a powerful mechanism for resource scoping, particularly when the range of permissible actions can be codified in a structured format.

\subsection{Controlled Natural Languages} While natural language permissions are flexible, they lack specificity. Controlled Natural Languages (CNLs) (i.e., subsets of natural language with restricted grammar and vocabulary), represent an interesting middle ground between structured and freeform specification. They preserve some of the readability of natural language while being more suitable for automated parsing and formal verification. An agent using a CNL interface might be able to interpret requests unambiguously, which reduces the risk of accidental misinterpretation. However, designing a CNL that is both secure and expressive can be challenging: allowing too much freedom increases ambiguity and exposes LLMs to prompt injection attacks \citep{promptInjection}, while a CNL that is too restricted will suffer from the same issues as structured languages.

\section{Example Use Cases}\label{sec:examples}

This section outlines four scenarios where authenticated delegation ensures secure and accountable AI agent interactions. Each example illustrates the structure of delegation credentials, the scoping mechanisms they enforce, and their role in maintaining accountability.

\subsection{AI Agent for Web Browsing}\label{sec:web-browsing}

\paragraph{Scenario.} 
A user employs an AI agent to perform tasks such as scheduling appointments, retrieving information, and managing online payments. The agent’s access must be restricted to specific websites, with clear limitations on the actions it can perform, such as transaction amounts.

\paragraph{Approach.}
\begin{enumerate}[topsep=3pt, itemsep=2pt]
    \item \textbf{Delegation Credential.} The credential specifies:
    \begin{itemize}[leftmargin=1em]
        \item \emph{User Identity:} The unique identity of the delegating user.
        \item \emph{Agent Identity:} A unique identifier for the agent, including its capabilities (e.g., browser-based interactions).
        \item \emph{Scope:} Restrictions such as approved websites, permitted actions (e.g., viewing schedules, making payments), and specific constraints (e.g., spending limits, validity duration).
    \end{itemize}
    \item \textbf{Access Enforcement.} Websites validate the agent’s credential upon login or transaction attempts. Unauthorized actions, such as accessing unapproved sites or exceeding predefined limits, are automatically blocked.
    \item \textbf{Auditability.} Logs tied to the agent’s unique identity record all transactions and actions, enabling post-interaction review and traceability.
\end{enumerate}

\paragraph{Why It Matters.} 
The structured credential ensures the agent cannot access unauthorized websites or perform unintended actions. This protects sensitive user data and ensures the user retains control over their online interactions.

\subsection{API-Only Data Manager}\label{sec:api-only}

\paragraph{Scenario.} 
An organization uses an AI agent to aggregate and analyze data from internal APIs, such as those providing information about operations or inventory. The agent’s access must be restricted to specific APIs and limited to non-destructive actions like querying data.

\paragraph{Approach.}
\begin{enumerate}[topsep=3pt, itemsep=2pt]
    \item \textbf{Delegation Credential.}
    \begin{itemize}[leftmargin=1em]
        \item \emph{User Identity:} The authenticated identity of the delegating organization or individual.
        \item \emph{Agent Identity:} A unique identifier for the agent, specifying its purpose (e.g., data aggregation).
        \item \emph{Scope:} Access permissions restricted to specific APIs, with limitations on actions (e.g., read-only access) and operational constraints (e.g., rate limits or expiration).
    \end{itemize}
    \item \textbf{API Enforcement.} APIs validate the credential and deny actions outside the granted permissions, such as attempts to write data or access restricted endpoints.
    \item \textbf{Credential Management.} Delegation tokens are periodically rotated or updated to reduce risks associated with stale credentials.
\end{enumerate}

\paragraph{Why It Matters.} 
The agent’s restricted scope ensures it cannot alter or access sensitive data unintentionally. Detailed access logs provide accountability and enable quick responses to anomalous behavior.

\subsection{Remote Virtual Environment via SSH}\label{sec:ssh-environment}

\paragraph{Scenario.} 
A user directs an AI agent to execute tasks in a remote virtual environment, such as running simulations or processing data. The agent’s access must be limited to specific commands and directories.

\paragraph{Approach.}
\begin{enumerate}[topsep=3pt, itemsep=2pt]
    \item \textbf{Delegation Credential.}
    \begin{itemize}[leftmargin=1em]
        \item \emph{User Identity:} The user’s authenticated identity with the virtual environment provider.
        \item \emph{Agent Identity:} A credential tied to the agent, specifying its role (e.g., simulation execution).
        \item \emph{Scope:} Permission to access specific directories, execute defined commands, and perform actions within a restricted time frame.
    \end{itemize}
    \item \textbf{Environment Enforcement.} The server enforces access control policies. Unauthorized actions, such as modifying configuration files or accessing sensitive directories, are rejected.
    \item \textbf{Audit Trail.} The environment logs each command executed by the agent, tied to its unique delegation credential, for post-task review.
\end{enumerate}

\paragraph{Why It Matters.} 
The restricted delegation credential ensures that the agent operates only within its assigned scope, safeguarding the environment against unintended or malicious actions.

\subsection{Agent-to-Agent Collaboration}\label{sec:agent-to-agent}

\paragraph{Scenario.} 
Two AI agents collaborate on a complex task, such as event planning or contract negotiation. Each agent has distinct roles and permissions that must be respected, such as one handling logistics and the other managing finances.

\paragraph{Approach.}
\begin{enumerate}[topsep=3pt, itemsep=2pt]
    \item \textbf{Delegation Credentials.}
    \begin{itemize}[leftmargin=1em]
        \item \emph{User Identity:} The authenticated identity of the delegating organization or individual.
        \item \emph{Agent Identities:} Each agent receives a unique credential describing its role and capabilities.
        \item \emph{Scopes:}
        \begin{itemize}[itemsep=1pt]
            \item Agent 1: Permissions for logistical tasks, such as booking services or scheduling.
            \item Agent 2: Permissions for financial tasks, such as processing payments, with explicit budget constraints.
        \end{itemize}
        \item \emph{Cross-Agent Verification:} Each agent includes its credential when issuing requests to the other. The receiving agent verifies the request is within scope before proceeding.
    \end{itemize}
    \item \textbf{Collaboration Mechanism.} The agents communicate using natural language, but all actionable requests reference their credentials for validation.
    \item \textbf{Auditability.} A log of all interactions, including credential references, ensures a clear record of tasks and decisions.
\end{enumerate}

\paragraph{Why It Matters.} 
By embedding scoping rules into cross-agent interactions, the collaboration remains secure and accountable. Each agent operates within its predefined limits, reducing the risk of unintended actions or miscommunications.

\end{document}